\documentclass{aastex}          
\usepackage{spr-astr-addons}   

\begin{document}
%
\title{Universe models with negative bulk viscosity}

\shorttitle{Universe Models with Negative Viscosity}
\shortauthors{Brevik \and Gr{\o}n}

\author{Iver Brevik\altaffilmark{1}}
\affil{Department of Energy and Process Engineering, Norwegian University of Science and Technology, Trondheim, Norway}
\email{iver.h.brevik@ntnu.no} 
\and
\author{{\O}yvind Gr{\o}n\altaffilmark{2}}
\affil{Oslo and Akershus University College of Applied Sciences, Faculty of Technology, Art and Design,  Oslo, Norway}
\email{oyvind.gron@hioa.no} 


\begin{abstract}
  The concept of negative temperatures has occasionally been used in connection with quantum systems. A  recent example of this sort is reported in the paper of S. Braun {\it et al.} [Science {\bf 339}, 52 (2013)], where an attractively interacting ensemble of ultracold atoms is investigated experimentally and found to correspond to a negative-temperature system since  the entropy decreases with increasing energy at the high end of the energy spectrum. As the authors suggest, it would be of interest to investigate whether a suitable generalization of standard cosmological theory could be helpful, in order to elucidate the observed accelerated expansion of the universe usually explained in terms of a positive tensile stress (negative pressure). In the present note we take up this basic idea and investigate a generalization of the standard viscous cosmological theory, not by admitting negative temperatures but instead by letting the bulk viscosity take {\it negative} values. Evidently, such an approach breaks standard thermodynamics, but may actually be regarded to lead to the same kind of bizarre  consequences  as the standard approach of admitting the equation-of-state parameter $w$  to be less than $-1$.  In universe  models dominated by negative viscosity we find that the fluid's entropy decreases with time,  as one would expect.  Moreover, we find that the fluid transition from the quintessence region into the phantom region (thus passing the phantom divide $w=-1$) can actually be reversed. Also in generalizations of the $\Lambda$CDM-universe models with a fluid having negative bulk viscosity we find that the viscosity decreases the expansion of the universe.
\end{abstract}


\section{Introduction}

The absolute temperature $T$ is in usual physics bound to be a positive quantity. Under special conditions, however, such as when high-energy states are more occupied than low-energy states, the temperature calculated from the thermodynamical formula
\begin{equation}
\frac{1}{T}=\left(\frac{\partial S}{\partial U}\right)_N, \label{1}
\end{equation}
can be a negative quantity. A striking example of this kind of system  has recently been found experimentally, in the form of  an attractively interacting ensemble of ultracold bosons; cf.  \cite{braun13}.

This is, however, not the first example of a negative-temperature system.  Thus negative temperatures are associated with the properties of paramagnetic dielectrics; cf. for instance, Ref.~\cite{landau80}, a key factor being here that the "magnetic spectrum" has to lie within a finite interval of energy. It is to be observed generally that the region of negative temperatures lies not "below absolute zero" but rather "above infinity", implying that negative temperatures are in some sense "higher" than positive ones.

An interesting idea suggested by \cite{braun13} is that the negative-temperature model may be helpful for the construction of a theory of dark energy in cosmology. As is commonly accepted by now, the expansion of the universe is a product of a  positive tensile stress, or negative pressure, that the cosmic fluid displays.

In which ways would it seem natural to generalize the standard cosmological theory? One possibility might be to allow for a composition of two, or more, components in the cosmic fluid. Thus the idea of considering the fluid as a mixture of two components, one ordinary fluid component and one dark energy component, has received substantial interest; cf., for instance, \cite{balakin11} and \cite{brevik12}. The line of approach in the present note will however be different, namely to allow for  {\it negative bulk viscosities} in the cosmic fluid. We shall consider the fluid as a one-component one. That means, we focus attention on the dark energy component only. In the real universe there may be a composition, as already mentioned, of a dark energy component and one or more other components, corresponding to normal fluids and perhaps also dark matter fluids.

At first sight the possibility $\zeta <0$ might seem unreasonable,  but one has to recall here that the positivity of $\zeta$ relies upon the physical requirement that the time development of entropy in a non-equilibrium system is a positive quantity. And that is just the condition that we wish to relax. Moreover, the characteristic property of the dark energy fluid implying that the parameter $w$ occurring in the equation of state,
\begin{equation}
p=w\rho, \label{2}
\end{equation}
is less than $-1$, is a counterintuitive property of the same kind. So, we think that the possibility of allowing for negative values of $\zeta$ is not so unreasonable after all, in view of the general bizarre properties of the dark energy fluid. As we will show in the next section, the reversal of the sign of $\zeta$  implies in a natural way that the entropy  of the fluid decreases with time. Thus we are discussing a viscosity-induced, rather than a temperature- induced, violation of the conventional second law in thermodynamics.

Further examples of universe models with negative bulk viscosity are investigated in sections 4 and 5. While positive viscosity accelerates the expansion of the universe, negative viscosity will in general decrease it.

In the summary section, section 6, we trace out some connections with other lines of approach investigated in contemporary cosmology. Whereas we keep $T$ positive and find the entropy change with respect to time to be negative, there are other approaches in which $T$ is negative and the corresponding entropy change positive. In some sense there is an equivalence.

\section{The formalism}

It is instructive to start from nonrelativistic theory. Let $u_i$ denote the components of the fluid velocity. The entropy density is $S=n\sigma$, where $n$ is the particle (baryon) density and $\sigma$ the entropy per particle (we use geometric units). Then, if $\eta$ denotes the shear viscosity and $\zeta$ as before the bulk viscosity, we have
\begin{equation}
\frac{dS}{dt}=\frac{2\eta}{T}(\theta_{ik}-\frac{1}{3}\delta_{ik} \nabla \cdot{\bf u})^2+\frac{\zeta}{T}(\nabla \cdot{\bf  u})^2+\frac{\kappa}{T}(\nabla T)^2, \label{3}
\end{equation}
where $\theta_{ik}=u_{(i; k)}$ and $\kappa$ is the thermal conductivity;  $\zeta$ (as well as $\eta$ and $\kappa$) are positive quantities in the conventional theory.

This expression can readily be generalized into a relativistic language. We here need some definitions. Let $U^\mu=(U^0, U^i)$ be the four-velocity of the fluid; in comoving coordinates $U^0=1, U^i=0$. If $g_{\mu\nu}$ is the general metric, the projection tensor is
\begin{equation}
h_{\mu\nu}=g_{\mu\nu}+U_\mu U_\nu, \label{4}
\end{equation}
the rotation tensor is
\begin{equation}
\omega_{\mu\nu}=h_\mu^\alpha h_\nu^\alpha U_{[\alpha; \beta]}, \label{5}
\end{equation}
the expansion tensor is
\begin{equation}
\theta_{\mu\nu}=h_\mu^\alpha h_\nu^\beta U_{(\alpha;\beta)}, \label{6}
\end{equation}
and finally the shear tensor is
\begin{equation}
\sigma_{\mu\nu}=\theta_{\mu\nu}-\frac{1}{3}h_{\mu\nu}\theta, \label{7}
\end{equation}
where $\theta=\theta_\mu^\mu={U^\mu}_{;\mu}$ is the scalar expansion.

With the spacelike heat flux density vector defined as
\begin{equation}
Q^\mu=-\kappa h^{\mu\nu}(T_{,\nu}+TA_\nu), \label{8}
\end{equation}
where $A_\nu=U^\alpha U_{\nu;\alpha}$ is the four-acceleration, we can now make the effective substitutions
\begin{equation}
\theta_{ik} \rightarrow \theta_{\mu\nu}, \quad \delta_{ik} \rightarrow h_{\mu\nu}, \quad \nabla \cdot {\bf u} \rightarrow \theta, \quad -\kappa T_{,k}\rightarrow Q_\mu, \label{9}
\end{equation}
whereby
\begin{equation}
{S^\mu}_{;\mu}=\frac{2\eta}{T}\sigma_{\mu\nu}\sigma^{\mu\nu}+\frac{\zeta}{T}\theta^2+\frac{1}{\kappa T^2}Q_\mu Q^\mu. \label{10}
\end{equation}
Here $S^\mu$ is the entropy current four-vector
\begin{equation}
S^\mu=n\sigma U^\mu+\frac{1}{T}Q^\mu. \label{11}
\end{equation}
(The sketchy derivation above follows \cite{brevik94}; more complete treatments can be found in \cite{weinberg71} and \cite{taub78}.)

Assume now spatial isotropy, implying $\eta=0$, and assume that there is no heat flux, $Q^\mu=0$. Then
\begin{equation}
{S^\mu}_{;\mu}=\frac{\zeta}{T}\theta^2. \label{12}
\end{equation}
We consider henceforth the  spatially flat FRW spacetime, with metric
\begin{equation}
ds^2=-dt^2+a^2(t)(dr^2+r^2d\Omega^2), \label{13}
\end{equation}
implying that $\theta=3H$ with $H$ the Hubble parameter. As ${S^\mu}_{;\mu}=n\dot{\sigma}$ in the local rest frame, we have
\begin{equation}
\dot{\sigma}=\frac{\zeta}{nT}\theta^2=\frac{9\zeta}{nT}H^2. \label{14}
\end{equation}
Thus, if $T$ still means the conventional positive temperature in the cosmic fluid, one gets $\dot{\sigma}<0$ if $\zeta <0$. The specific entropy $\sigma$ has to decrease  with increasing time in this model.

\section{Remarks on the phantom divide}

A more detailed insight into the physics of this model can be achieved by considering the behavior near the phantom divide more closely. This divide is defined as the case $w=-1$. As is known from observations \cite{nakamura10,amanullah10},  $w$ lies close to $-1$ today:
\begin{equation}
w=-1.04^{+0.09}_{-0.10}. \label{15}
\end{equation}
A characteristic property of most of the phantom dark energy models is the occurrence of the Big Rip future singularity: once the phantom divide is crossed so as to give $w<-1$, the universe is inevitably driven into a singularity (the scale factor becoming infinity) at a finite time in the future. This was first observed by  \cite{caldwell02}, and has later been re-examined by a number of researchers (for a recent review  including also modified gravity, see, for instance, \cite{nojiri11}).

These early theories assumed the cosmic fluid to be nonviscous. Once bulk viscosity is included, the theory becomes richer and more flexible. One important property, on which we shall focus attention in the following, is that on the basis of a conventional positive value of $\zeta$ it becomes possible for the fluid to slide from the quintessence region ($-1<w<-1/3$) through the phantom divide into the phantom region and thus afterwards into the future singularity. This was first pointed out in \cite{brevik05}. Whether a transition through the "point" $w=-1$ really occurs or not, depends on the magnitude of $\zeta$.

Consider now the Friedmann equations for the flat space,
\begin{equation}
\theta^2= 3\kappa  \rho, \label{16}
\end{equation}
\begin{equation}
2\dot{\theta}+\theta^2=-3\kappa(p-\zeta \rho). \label{17}
\end{equation}
Together with the conservation equation for energy,
\begin{equation}
\dot{\rho}+(\rho +p)\theta=\zeta \theta^2, \label{18}
\end{equation}
they provide a set of equations enabling us to derive the governing equation for the scalar expansion, or equivalently, for the energy density. Imagine first the general case for which $w=w(\rho)$. Then, if the function $f(\rho)$ is defined via
\begin{equation}
1+w(\rho)=-f(\rho)/\rho, \label{19}
\end{equation}
we can write the governing equation for $\rho$ in the form
\begin{equation}
\dot{\rho}-\sqrt{3\kappa \rho}\, f(\rho)-3\kappa \rho \zeta(\rho)=0, \label{20}
\end{equation}
which has the solution
\begin{equation}
t=\frac{1}{\sqrt {3\kappa}}\int_{\rho_0}^\rho \frac{d\rho}{\sqrt{\rho}f(\rho)[1+\sqrt{3\kappa}\, \zeta(\rho)\sqrt{\rho}/f(\rho)]}. \label{21}
\end{equation}
Here we have taken $t=t_0=0$ as the initial point, $\rho_0$ meaning $\rho(t_0)$.

We limit ourselves in the following to the case when $f(\rho)=\alpha \rho$, with $\alpha$ a constant. Thus
\begin{equation}
p=w\rho=-(1+\alpha)\rho. \label{22}
\end{equation}
We next need to model the form of the bulk viscosity. Probably the most interesting form from a physical point of view is to put the viscosity proportional to the scalar expansion. Therewith we allow for an increase of the viscosity in the case of increasingly vigorous movements in the cosmic fluid. Let us assume that
\begin{equation}
\zeta =\tau \theta, \label{23}
\end{equation}
with $\tau$ a constant. This choice has been analyzed repeatedly also before; cf., for instance, \cite{brevik05,gron90,brevik13}. Equation (\ref{21}) then yields
\begin{equation}
t=\frac{1}{\sqrt{3\kappa}}\,\frac{2}{\alpha+3\kappa \tau}\left(\frac{1}{\sqrt {\rho_0}}-\frac{1}{\sqrt \rho}\right). \label{24}
\end{equation}
This shows that the fate of the universe is critically dependent of the sign of the prefactor. The condition for a Big Rip $(\rho =\infty)$ to occur, is that
\begin{equation}
\alpha+3\kappa \tau>0. \label{25}
\end{equation}
In conventional viscous cosmology, as explored in \cite{brevik05}, even if the universe starts from a state lying in the quintessence region $(\alpha <0)$, it is possible for the fluid to slide through the phantom divide into the phantom region if the viscosity is big enough $(\tau>0$ in this case). Once having entered the phantom region, the Big Rip in one of its variants becomes inevitable.

In the present case, however, with $\tau<0$, the situation becomes reversed. Even if the fluid starts within the phantom region $(\alpha>0)$ it is possible, if the negative viscosity becomes large enough in magnitude, to abandon the singularity by making the expression on the left hand side of (\ref{25}) negative. The fluid goes back to the quintessence region $w>-1$, and becomes thus infinitely thinned in the far future, $\rho \rightarrow 0$. The role of the phantom divide as a kind of a one-way "membrane" is in this way no longer upheld.

\section{Generalization of the $\Lambda$CDM universe model with negative bulk viscosity}

We here use the results of \cite{mostafapoor11}, but allow for the possibility that the bulk viscosity is negative. Let us first consider a flat universe model with dust, and LIVE in which the interaction of the dust and the vacuum energy with stress and thus with negative absolute temperature, is modeled by negative viscosity. The Raychaudhury equation may then be written
\begin{equation}
\dot{H}+\frac{3}{2}H^2-\frac{3}{2}\kappa \zeta H- \frac{\kappa}{2}\rho_\Lambda=0, \label{26}
\end{equation}
where $\kappa=8\pi G$ is Einstein's gravitational constant and $\rho_\Lambda$ is the density of the Lorentz Invariant Vacuum Energy, LIVE, which is constant and may be represented by the cosmological constant.

We shall first consider the case where the bulk viscosity is constant and negative, $\zeta=\zeta_0<0$. Then the general solution of this equation with $a(0)=0$ and $a(t_0)=1$  may be written
\begin{subequations}\label{27}
\begin{eqnarray}
  H(t)&=&\frac{\kappa}{2}\zeta_0+\alpha \coth \left(\frac{3}{2}\alpha t\right), \\
  \alpha&=&\left( \frac{1}{4}\kappa^2\zeta_0^2+\frac{1}{3}\kappa \rho_\Lambda \right)^{1/2}
\end{eqnarray}
\end{subequations}
\begin{subequations}\label{28}
\begin{eqnarray}
a(t)&=&\beta \exp{ \left[\frac{\kappa \zeta_0}{2}(t-t_0)\right]}\sinh^{\frac{2}{3}} \left( \frac{3}{2}\alpha t\right),\\
\beta&=&\left[\frac{\rho_{M0}-3H_0\zeta_0}{\rho_\Lambda+(3/4)\kappa \zeta_0^2}\right]^{1/3},
\end{eqnarray}
\end{subequations}
where $\rho_{M0}$ is the present density of cold dark energy which has been assumed to be pressure less dust, and $H_0=H(t_0)$  is the present value of the Hubble parameter.

Note that the value of $\alpha$ is independent of the sign of $\zeta_0$. The Hubble parameter has an infinitely large initial value and decreases towards $H_\infty=\kappa \zeta_0/2+\alpha$, which is positive for all values of $\zeta_0$.
  It is seen that when $\zeta_0$ is constant the sign of $\zeta_0$ does change the behavior of the universe qualitatively. The age of the universe is
\begin{equation}
t_0=(2/3\alpha){\rm arcsinh} (1/\beta)^{3/2}, \label{29}
\end{equation}
showing that negative viscosity makes the age smaller.

We shall then consider the case where the bulk viscosity is proportional to the Hubble parameter with a negative constant of proportionality, $\zeta=\zeta_1 H, \zeta_1<0$. In this case the Raychaudhuri equation takes the form
\begin{equation}
\dot{H}+\frac{3}{2}(1-\kappa \zeta_1)H^2 -\frac{\kappa}{2}\rho_\Lambda=0. \label{30}
\end{equation}
Solving this equation with the boundary condition that $H(t_0)=H_0$ gives
\begin{subequations}\label{31}
\begin{eqnarray}
H(t)&=&H_1\coth \left(\frac{3}{2}H_2t\right), \\
H_1&=&H_0\sqrt{\frac{\Omega_{\Lambda0}}{1-\kappa \zeta_1}}, \\
H_2&=&H_0\sqrt{(1-\kappa \zeta_1)\Omega_{\Lambda 0}}.
\end{eqnarray}
\end{subequations}
The scale factor is
\begin{equation}
a(t)=\left[ \frac{\sinh \left(\frac{3}{2}H_2t\right)}{\sinh \left( \frac{3}{2}H_2t_0\right)}\right]^{\frac{2}{3(1-\kappa \zeta_1)}}. \label{32}
\end{equation}
The age of this universe model is
\begin{equation}
t_0=\frac{2}{3H_0}\frac{1}{\sqrt{\Omega_{\Lambda 0}(1-\kappa \zeta_1)}}
{\rm arccoth} \sqrt{\frac{1-\kappa \xi_1}{\Omega_{\Lambda 0}}}. \label{33}
\end{equation}
In the present case the age-redshift relationship, i.e. the relationship between the time $t$ of emission and the time $t_0$ of observation of radiation with redshift $z$ is
\begin{equation}
t=t_0\frac{{\rm arcsinh}\left[ \frac{\sqrt{\Omega_{\Lambda 0}}}{(1+z)^{\frac{3}{2}(1-\kappa \zeta_1)}\sqrt{1-\kappa \zeta_1-\Omega_{\Lambda0}} }\right]}
{{\rm arctanh} \sqrt{\frac{\Omega_{\Lambda 0}}{1-\kappa \zeta_1}}}. \label{33A}
\end{equation}
The corresponding expression for the standard universe model without viscosity \cite{gron02} is obtained by putting $\zeta_1=0.$

For this universe model the equation of continuity takes the form
\begin{equation}
\dot{\rho}_M+3(1-\kappa \zeta_1)H\rho_M-3\kappa \zeta_1\rho_\Lambda H=0. \label{34}
\end{equation}
Inserting the expression (\ref{31}) for $H$ we find that Eq.~(\ref{34}) has the general solution
\[
\rho_M(t)=\frac{c_1}{{\rm sinh}^2\left(\frac{3}{2}H_2t\right)}+\rho_{M\infty}, \quad \rho_{M\infty}=\frac{\kappa \zeta_1\rho_\Lambda}{1-\kappa \zeta_1}, \]
\begin{equation}
 \quad c_1=(\rho_{M0}-\rho_{M\infty}){\rm sinh}^2\left(\frac{3}{2}H_2t_0\right), \label{35}
\end{equation}
where $\rho_{M0}=\rho(0)$ and $\rho_{M\infty}=\lim|_{t\rightarrow \infty}\rho(t)$. Note that $\rho_{M\infty}<0$ for $\zeta_1<0$. Hence there exists an instant $t_1$ where the density of the dust vanishes,  after which this model is no more physically realistic. The equation  $\rho(t_1)=0$ leads to
\begin{equation}
{\rm sinh}\left(\frac{3}{2}H_2t_1\right)=
\sqrt{1-\frac{\rho_{M0}}{\rho_{M\infty}}}
{\rm sinh}\left(\frac{3}{2}H_2t_0\right). \label{36}
\end{equation}
The scale factor and the Hubble parameter at this instant are
\begin{equation}
a(t_1)=\left( 1-\frac{\rho_{M0}}{\rho_{M\infty}}\right)^{\frac{1}{3(1-\kappa \zeta_1)}}, \label{37}
\end{equation}
\begin{equation}
H(t_1)=\left( \frac{H_1^2\rho_{M0}-H_0^2\rho_{M\infty}}{\rho_{M0}-\rho_{M\infty}}\right)^{1/2}. \label{38}
\end{equation}
Here there is no singularity at the instant $t_1$, only a transition to an unphysical state with negative mass density.

\section{How a sign-change of viscosity modifies the properties of some viscous universe models}

We shall here consider some viscous universe models that have been studied earlier with positive bulk viscosity, and investigate how their physical properties become changed when the viscosity changes sign.

If the vacuum energy is removed in the universe model described by Eqs.~(\ref{27}) and (\ref{28}), and the fluid is assumed to obey an equation of state $p=w\rho$ with constant value of $w$, the scale factor with $a(0)=0, a(t_0)=1$ is given by \cite{treciokas71}
\begin{subequations}\label{39}
\begin{eqnarray}
a^{\frac{3}{2}(1+w)}&=&\frac{2}{3\zeta_0}\left( \exp{(\frac{3}{2}\zeta_0t)}-1\right), \\
t_0&=&\frac{2}{3\zeta_0}\ln \left( 1+\frac{3}{2}\zeta_0\right). \label{39a}
\end{eqnarray}
\end{subequations}
Hence with $\zeta_0>0$ this universe model will eventually reach a viscosity dominated steady state era with exponential expansion and $\lim a|_{t\rightarrow \infty}=\infty$, but when $\zeta_0<0$ the effect of the viscosity upon the expansion of the universe diminishes, and the universe model has a finite final value of the scale factor, $\lim a|_{t\rightarrow \infty}=-2/3\zeta_0$.

Some years ago, \cite{murphy73} considered a class of viscous universe models dominated by a viscous fluid with $p=w\rho$ and bulk viscosity $\zeta=\gamma \rho$ where $\gamma$ is constant. Then the rate of change of the Hubble parameter is
\begin{equation}
\dot{H}=\frac{3}{2}H^2(3\gamma H-1-w). \label{40}
\end{equation}
The solution of this equation with $a(t_0)=1, H(t_0)=H_0$ is given by the equation
\begin{subequations}\label{41}
\begin{eqnarray}
\frac{2\gamma}{1+w}\ln \left(3\gamma -\frac{1+w}{H}\right) &+&\frac{2}{3}\left(\frac{1}{H}-\frac{1}{H_0}\right) \nonumber\\
&=&(1+w)(t-t_0), \\
H_0&=&\frac{1+w}{3\gamma-1}.
\end{eqnarray}
\end{subequations}
On the other hand inserting $H=\dot{a}/a$ in the last factor of Eq.~(\ref{40}) and then integrating with the same boundary conditions we obtain
\begin{equation}
3\gamma \ln a+\frac{2}{3}\left( \frac{1}{H}-\frac{1}{H_0}\right)=(1+w)(t-t_0). \label{42}
\end{equation}
Comparison between Eqs.~(\ref{41}) and (\ref{42}) gives
\begin{equation}
a^{\frac{3}{2}(1+w)}=3\gamma-\frac{1+w}{H}. \label{43}
\end{equation}
For positive viscosity ($\gamma>0$) these equations describe an expanding universe  model, but for negative viscosity ($\gamma<0$) the Hubble parameter must be negative for the scale factor to be positive. Hence the universe model contracts.

As an example of anisotropic universe models we shall finally consider the effect of negative viscosity of some simple universe models of Bianchi type I. The line element has the form
\begin{equation}
ds^2=-dt^2+\sum_{i=1}^3 a_i^2(t)(dx^i)^2. \label{44}
\end{equation}
The directional Hubble parameters are $H_i=\dot{a}_i/a_i$, and the anisotropy parameter is
\begin{subequations}\label{45}
\begin{eqnarray}
  A&=&\frac{1}{3}\sum_{i=1}^3 \left( \frac{\Delta H_i}{H}\right)^2, \\
   \Delta H_i&=&H_i-H, \quad H=\frac{1}{3}(H_1+H_2+H_3).
\end{eqnarray}
\end{subequations}
In a universe filled with LIVE with constant density $\rho_\Lambda$ we have;  cf. \citet{mostafapoor13}
\begin{equation}
\kappa (\rho+\rho_\Lambda)=\frac{3}{2}(2-A)H^2. \label{46}
\end{equation}
If $\rho_\Lambda=0$ and in the case of a relativistically rigid fluid with $w=1$ (a so-called Zel'dovich fluid), with constant bulk viscosity $\zeta$, the anisotropy parameter as a function of time is
\begin{equation}
A=A_0e^{-3\zeta (t-t_0)}. \label{47}
\end{equation}
Hence, a positive viscosity leads to decay of the anisotropy, while a negative viscosity increases the anisotropy. This is a characteristic behavior for more general viscous anisotropic universe models. Including LIVE the Hubble parameter is
\begin{equation}
H(t)=\frac{\kappa \zeta}{4}+\hat{H}\coth (3\hat{H}t), \quad \hat{H}^2=\left( \frac{\kappa \zeta}{4}\right)^2+\frac{\kappa \rho_\Lambda}{3}. \label{48}
\end{equation}
It is seen that the sign of the viscosity does not influence $\hat H$, but the average Hubble factor $H$ is smaller with negative viscosity than with a positive one. Negative viscosity acts like a decelerating force upon the expansion of the universe.

\section{Summary}
Our purpose in this paper has been to investigate situations in cosmology where the entropy decreases with increasing time. Specifically, we have achieved this by taking the bulk viscosity $\zeta$ to be less than zero (in accordance with spatial isotropy, the shear viscosity has been put equal to zero). The ansatz $\zeta <0$  is of course  counterintuitive, but one should here note that the positivity of $\zeta$ in conventional cosmology is based upon the requirement that the change of entropy in a non-equilibrium system is positive, and that is just the property that we wish to relax. One may also observe the analogy with the phantom era in the expansion of the universe, meaning that the parameter $w$ in the equation of state $p=w\rho$ is less that $-1$. In both cases, bizarre thermodynamic behaviors are encountered.

We have shown that in a generalization of the $\Lambda$CDM universe model with negative bulk viscosity, the viscosity contributes with an attractive gravity, and hence tends to decrease the expansion. It turns out that in a model where the negative coefficient of bulk viscosity is proportional to the density of the fluid, expansion is not allowed. Therefore, after all,  even if negative viscosity is a theoretical possibility, it does not seem to be a favored property of the cosmic fluid.

Finally, it is of interest to put our developments into a wider perspective by comparing them with some other works in modern cosmology.

$\bullet$ Our formalism allows for the presence of a negative bulk viscosity $\zeta$ but keeps the temperature $T$ positive. We emphasize that the entropy four-vector (\ref{11}) does not depend upon the sign of $\zeta$ at all. What changes sign with $\zeta$, is the {\it change} of entropy with time; cf. Eq.~(\ref{10}).

$\bullet$ In many cases, the inclusion of bulk viscosity in cosmological theory does not lead to significant changes. For instance, the Cardy-Verlinde formula for entropy, cf.  \cite{verlinde00}, has been found to apply under various conditions in the presence of viscosity, even in the case of a multicomponent fluid obeying an inhomogeneous equation of state; cf.  \cite{brevik02,brevik02a,brevik10}. A somewhat stronger influence from cosmology is experienced, as mentioned above, in cases where the viscosity is large enough to make the fluid pass through the phantom barrier $w=-1$ into the phantom regime (\cite{brevik05}).

$\bullet$ In other cases, when dealing with dark energy, one expects that the the entropy itself can be negative. Thus, \cite{nojiri05} considered the effect of a dark energy ideal fluid by inserting an inhomogeneous Hubble-parameter dependent term in the late-time universe. Remarkably enough, a thermodynamical dark energy model was found in which, despite preliminary expectations  (\cite{brevik04}), the entropy of the phantom epoch could be positive. This was caused by crossing of the phantom barrier. Theories of this kind are wider in scope, and generally different from, the one presented by us above.

\end{document}